\documentclass[aps,prl,letterpaper,twocolumn,showpacs,superscriptaddress]{revtex4}
\usepackage{longtable}
\usepackage{graphicx}
\usepackage{epsfig}
\usepackage{amssymb}
\usepackage{color} 
\usepackage{sidecap}

\newcommand{\coon}{(Color online) }

\begin{document}

\title{Coherent control of  vacuum squeezing in the Gravitational-Wave Detection Band}

\author{Henning Vahlbruch}
\affiliation{Max-Planck-Institut f\"ur Gravitationsphysik (Albert-Einstein-Institut) and\\ Institut f\"ur Gravitationsphysik, Universit\"at Hannover, Callinstr. 38, 30167 Hannover, Germany}

\author{Simon Chelkowski}
\affiliation{Max-Planck-Institut f\"ur Gravitationsphysik (Albert-Einstein-Institut) and\\ Institut f\"ur Gravitationsphysik, Universit\"at Hannover, Callinstr. 38, 30167 Hannover, Germany}

\author{Boris Hage}
\affiliation{Max-Planck-Institut f\"ur Gravitationsphysik (Albert-Einstein-Institut) and\\ Institut f\"ur Gravitationsphysik, Universit\"at Hannover, Callinstr. 38, 30167 Hannover, Germany}

\author{Alexander Franzen}
\affiliation{Max-Planck-Institut f\"ur Gravitationsphysik (Albert-Einstein-Institut) and\\ Institut f\"ur Gravitationsphysik, Universit\"at Hannover, Callinstr. 38, 30167 Hannover, Germany}

\author{Karsten Danzmann}
\affiliation{Max-Planck-Institut f\"ur Gravitationsphysik (Albert-Einstein-Institut) and\\ Institut f\"ur Gravitationsphysik, Universit\"at Hannover, Callinstr. 38, 30167 Hannover, Germany}

\author{Roman Schnabel}
\affiliation{Max-Planck-Institut f\"ur Gravitationsphysik (Albert-Einstein-Institut) and\\ Institut f\"ur Gravitationsphysik, Universit\"at Hannover, Callinstr. 38, 30167 Hannover, Germany}

\date{\today}

\begin{abstract}
We propose and demonstrate a coherent control scheme for stable phase locking of squeezed vacuum fields. We focus on sideband fields at frequencies from 10\,Hz to 10\,kHz which is a frequency regime of particular interest in gravitational wave detection and for which conventional control schemes have failed so far.  A vacuum field with broadband squeezing covering this entire band was produced using optical parametric oscillation and characterized with balanced homodyne detection. The system was stably controlled over long periods utilizing two coherent but frequency shifted control fields. In order to demonstrate the performance of our setup the squeezed field was used for a nonclassical sensitivity improvement of a Michelson interferometer at audio frequencies.
\end{abstract}

\pacs{04.80.Nn, 42.50.Lc, 42.65.Yj, 95.55.Ym}

\maketitle

It was first proposed by Caves~\cite{Cav81} that injected squeezed states may be used to improve the sensitivity of laser interferometers and might therefore contribute to the challenging effort of direct observation of gravitational waves \cite{Thorne87}. The goal of that proposal was the reduction of the measurement's shot noise. Later Unruh~\cite{Unruh82} has found that squeezed light can be used to correlate interferometer shot noise and radiation pressure noise thereby breaking the so-called standard quantum limit and allowing for a quantum nondemolition measurement on the mirror test mass position, for an overview we refer to Ref.\,\cite{KLMTV01}. Harms {\it et al.}\,\cite{HCCFVDS03} have shown that advanced interferometer recycling techniques \cite{Mee88} that also aim for an improvement of the signal-to-shot-noise-ratio are fully compatible with squeezed field injection.
%
%
Gravitational wave detectors require squeezing in their detection band from about 10\,Hz to 10\,kHz. 
The majority of current squeezing experiments, however, have been performed in the MHz regime.  Furthermore the orientation of the squeezing ellipse needs to be designed for every sideband frequency. The transformation from frequency independent squeezing to optimized frequency dependent squeezing can be performed by optical filter cavities as proposed in \cite{KLMTV01} and demonstrated in \cite{CVHFLDS05} for MHz frequencies. Also in the MHz-regime, the combination of squeezed field injection and recycling techniques has been demonstrated \cite{KSMBL02,VCHFDS05}. Squeezing at audio frequencies has been demonstrated recently for the first time \cite{MGBWGML04,MGGLM06}. However, the phase of the squeezed vacuum could not be controlled by a coherent field.

 Controlling squeezed vacuum fields is the basic problem for squeezed field applications in GW detectors. Common control schemes rely on the injection of a weak, phase modulated seed field at the carrier frequency into the OPO  thereby turning the device into an optical parametric amplifier (OPA). It has been shown that even lowest carrier powers introduce large amounts of classical laser noise at audio frequencies and squeezing can no longer be achieved \cite{MGBWGML04}. On the other hand phase modulation sidebands are not present in a pure vacuum field. 
For this reason  in \cite{MGBWGML04,MGGLM06} a coherent control field for locking the squeezed quadrature angle to a local oscillator could not be created. The quadrature angle was locked instead using so-called noise locking whose stability was found to be significantly less than what can be achieved with coherent modulation locking \cite{MMGLGGMM05} as used in GW interferometers.
 
In this Letter we report on the demonstration of a coherent control scheme for stable phase locking of squeezed vacuum fields. The scheme was used to produce broadband squeezing at audio and sub-audio frequencies covering the complete detection band of ground based GW detectors.

\begin{figure*}[t]
  \hspace{-80mm}
\includegraphics[width=9cm]{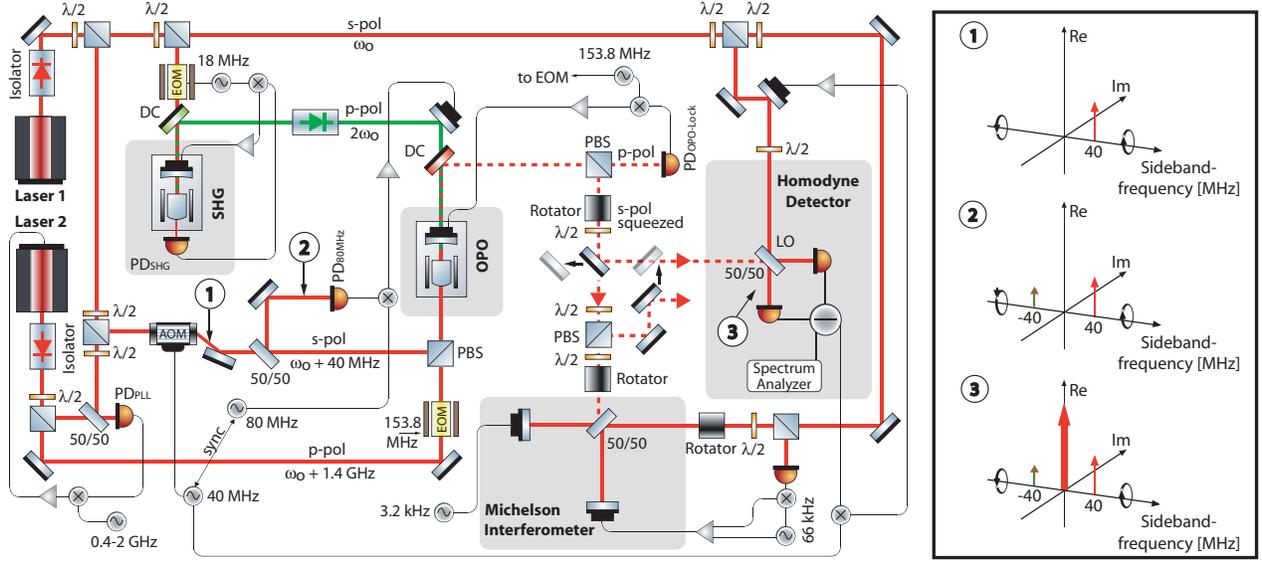}
\caption{\coon  Left: Schematic of the experiment. Generation and full coherent control of a broadband squeezed vacuum field at 1064\,nm was achieved utilizing two independent but phase locked laser sources. Laser 1 provided the main carrier frequency of homodyning local oscillator and Michelson interferometer ($\omega_{0}$). It also provided a frequency shifted control field utilizing an acousto-optical-modulator (AOM) and the optical parametric oscillator (OPO) pump field after second harmonic generation (SHG). Laser 2 provided another frequency shifted control field. 
PBS: polarizing beamsplitter;  DC: dichroic mirror; LO: local oscillator, PD: photo diode; EOM: electro-optical modulator, \vrule width 3mm depth -.6mm\,: piezo-electric transducer. 
Right: Complex optical field amplitudes at three different locations in the experiment.
}
  \label{experiment}
\end{figure*}

The general purpose of an interferometer is to transform an optical phase modulation signal into an amplitude modulation which can be measured by a single photo diode or by a balanced homodyne detector. The observable is described by the time-dependent quadrature operator $\hat q_\theta (\Omega, \Delta\Omega, t)$ where $\Omega$ denotes the modulation sideband frequency, $\Delta\Omega$ the detection resolution bandwidth (RBW) and $\theta$ the quadrature angle. 
The amplitude quadrature ($\theta\!=\!0$) is usually denoted by subscript 1 and phase quadrature ($\theta\!=\!\pi/2$) by subscript 2 \cite{CVHFLDS05}. For the vacuum state the variances of quadrature operators ($\Delta \! ^{2} \hat q_\theta$) are normalized to unity and, generally,  Heisenberg's Uncertainty principle sets a lower bound for the product of the two variances of non-commuting pairs of quadrature operators, e.g. 
 $ \Delta \!  ^{2} \hat q_1(\Omega, \Delta\Omega, t) \cdot \Delta \!  ^{2} \hat q_2(\Omega, \Delta\Omega, t) \ge 1. $
If an ideal phase-sensitive amplifier such as a loss-less OPO acts on the vacuum state the equality  still holds but one quadrature variance is squeezed. For a nonclassical improvement of an interferometer's signal-to-noise-ratio the angle of the squeezed quadrature then needs to be aligned to the modulation signal of interest.
In realistic situations the degree of squeezing is decreased by optical losses thereby mixing the squeezed state with a vacuum, as well as by noisy classical modulation fields that beat with the same local oscillator, and also by acousto-mechanical disturbances during the squeezed field generation and during its measurement \cite{MGBWGML04}.

Noisy modulation fields from the laser source can be completely removed if the squeezed light source is just seeded by a pure vacuum field. To prevent any contamination of the squeezed field, an appropriate control scheme may solely use some additional fields that do not interfere with the squeezed mode. If nevertheless such control fields are coherent with the squeezed mode full control capability is possible.  The length of the OPO might be controlled by using phase modulation sidebands on a different (frequency shifted) spatial mode or polarization mode. In the experiment presented here, we used the latter. 
However, control of the quadrature angle in respect to a homodyning local oscillator or an interferometer is more challenging. Here we propose to use another control field that does sense the OPO nonlinearity but is frequency detuned against the vacuum squeezed mode. When coupled into the OPO the parametric gain $g$ turns the single sideband field into the following field with amplified and deamplified quadratures, respectively:
\begin{eqnarray}
\label{lock}
E(t)  \approx e^{-i\omega_{0}t} \alpha_{\Omega} \left(\sqrt{g} \cos (\Omega t) + i \sin (\Omega t) /\sqrt{g} \right)\,+ c.c. \nonumber \,,
\end{eqnarray}
where $\alpha_{\Omega}$ is the complex amplitude of the single sideband field with carrier detuning $\Omega$ and with phase set to zero. When $E(t)$ is directly detected with a photo diode and demodulated at $2\Omega$ an error signal for the quadrature angle can be produced that allows coherent locking of the second harmonic pump field with respect to the control field. Furthermore, when $E(t)$ is overlapped with a local oscillator at carrier frequency $\omega_{0}$ and the photo current is demodulated at $\Omega$ another error signal can be derived to lock the control field to the local oscillator (LO) thereby locking the squeezed vacuum field to the LO. 

\begin{figure*}[t!]
  \hspace{-110mm}
  \includegraphics[width=5.5cm]{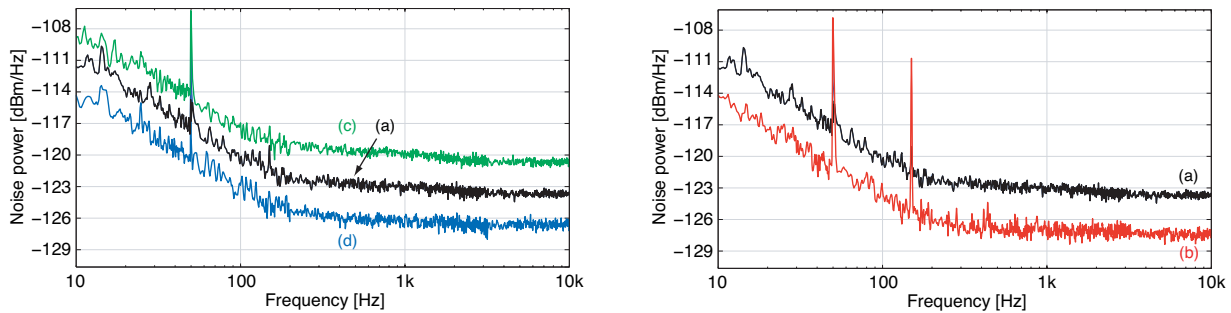}
  \caption{(color online). Measured quantum noise spectra. Left: Shot noise for three different local oscillator (LO) powers, (a) $88 \mu W$, (c) $176 \mu W$, and (d) $44 \mu W$. Right: (a) Shot noise and (b) squeezed noise with $88 \mu W$ LO power. All traces are pieced together from five FFT frequency windows: 10 Hz-50 Hz, 50 Hz-200 Hz, 200 Hz-800 Hz, 800 Hz-3.2 kHz, and 3.2 kHz-10 kHz. 
Each point is the averaged rms value of 100, 100, 400, 400 and 800
measurements in the respective ranges. The RBWs of the
five windows were 250\,mHz, 1\,Hz, 2\,Hz, 4\,Hz and 16\,Hz, respectively. Linearity of the detection system was carefully checked in a separate measurement.}
  \label{ShotcombiSQZ}
\end{figure*}

\begin{figure}[b]
\hspace{-20mm}
  \includegraphics[width=5.0cm]{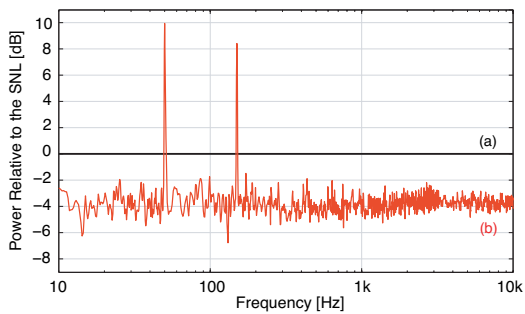}
  \vspace{0mm}
  \caption{(color online). Squeezed quantum noise plotted relatively to the shot-noise level. 
  Nonclassical noise suppression of about 4\,dB with an almost white spectrum is demonstrated.
  The measurement corresponds to Fig.\,2 (right).}
  \label{SQZdivSN}
\end{figure}

A schematic of the experiment is shown in Fig.\,\ref{experiment}. The main laser source (laser 1) was a monolithic non-planar Nd:YAG ring laser of 2\,W single mode output power at 1064\,nm. Approximately  1.4\,W was used to pump a second harmonic generation (SHG) cavity. The design of the SHG cavity was similar to the one of the OPO cavity described below but with an outcoupling mirror reflectivity of R=92\%. A more detailed description of our SHG can be found in \cite{CVHFLDS05}. 
About 60\,mW of the second harmonic field was effectively used to pump the squeezed light source. 
The OPO was an hemilithic optical resonator consisting of a  7\% doped MgO:LiNbO$_{3}$ crystal and an piezo mounted  output coupler. The power reflectivity of the coupling mirror was 95.6\% at 1064\,nm and 20\% at 532\,nm. The curved back surface of the crystal had a high reflection coating (R=99.96\%) for both wavelengths. The cavity's free spectral range was 3.8\,GHz. Orientation of the nonlinear crystal was such that s-polarized fields could sense parametric gain. For OPO length control we used a phasemodulated p-polarized field which was supplied by another monolithic non-planar Nd:YAG ring laser (laser 2). We determined a frequency shift of about 1.4\,GHz between the two polarization modes for simultaneous resonance inside the OPO. The frequency offset was controlled  via a high bandwidth phase locking loop. The p-polarized locking field was injected through the crystal's back surface and spatially separated from the s-polarized squeezed vacuum with a polarizing beam splitter (PBS) behind the OPO. An error signal for the OPO cavity length could be generated by demodulating the detected  p-polarized beam at a sideband frequency of 153.8\,MHz without introducing any significant loss on the squeezed field. We measured the loss to be less than 0.1\% given by the PBS's power reflectivity for s-polarized light  of 99.9\%. We point out that the p-polarized control field did not add any significant noise to the squeezed mode, given the low residual power contribution after the PBS and the high frequency offset.

Phase control with respect to a local oscillator at the fundamental frequency $\omega_{0}$ is required for both characterization and application of audio-band squeezed vacuum fields.
Following our proposal described above we utilized another coherent but frequency shifted control field. 
Detuned by 40\,MHz with respect to the main carrier frequency ($\omega_{0}$, laser 1) by an acousto-optical-modulator (AOM) this s-polarized infra-red field (440\,$\mu$W) was also injected into the OPO cavity where its quadratures are parametrically amplified and deamplified which generates a sideband at -40\,MHz.
This mechanism derived an error signal for controlling the phase relation between the 40\,MHz control field and the green pump. The error signal could be obtained by detecting the control field back-reflected from the OPO and demodulating the photo current at twice the beat frequency (80\,MHz) as illustrated in the sideband scheme in Fig.\,\ref{experiment}. By feeding back the error signal to an PZT-mounted mirror in the path of the green pump field stable phase control was realized. A separate locking loop was required to control the phase between the squeezed vacuum field, given by the phase of the green pump, and the LO. The appropriate error signal was derived at the homodyne detector, described in \cite{CVHFLDS05}. By taking the difference of both homodyne photo-diodes a demodulation at  a frequency of 40\,MHz provides an error signal which again was fed back to a phaseshifting PZT-mounted mirror.  Generated and detected fields at three  locations of the experiment are shown in the box in Fig.\,\ref{experiment}.  

Observation and characterization of squeezed fields at audio frequencies is usually performed with balanced homodyne detection as it is done in the MHz regime. However, at low frequencies measurement time increases and stable control of all the field's degrees of freedom is essential. In addition homodyne detection at audio frequencies requires a  much greater classical noise suppression.
Fig.\,\ref{ShotcombiSQZ}\,(left) verifies that our detector was indeed quantum  noise limited over the full spectrum from 10 Hz to 10 kHz. For these measurements the (squeezed) signal input in  Fig.\,\ref{experiment} was blocked. A change of local oscillator power by a factor of two resulted in a change of measured quadrature noise variance of the same factor, whereas classical noise from the local oscillator would scale with the squared factor. The nominal local oscillator power used was 88\,$\mu$W. In all measured spectra shown here the electronic noise of the detection system was at least 7\,dB below shot noise and has been subtracted from all data. 
The roll-up at low frequencies clearly corresponds to the transfer function of our homodyne detector. This transfer function was independently measured by modulating the local oscillator power at frequencies swept over the spectrum of interest.  

The OPO squeezing spectrum from 10\,Hz - 10\,kHz is shown in Fig.\,\ref{ShotcombiSQZ}\,(right). Trace (a) shows the shot noise limit of the homodyne detetction system whereas the squeezed quantum noise is shown in trace (b). The measurement time was approximately 1.5 hours during which the setup was stably controlled in all degrees of freedom.
Fig.\,\ref{SQZdivSN} clearly shows that 4\,dB squeezing has been achieved over the complete detection band of ground based GW interferometers. 
For this plot the traces from Fig.\,\ref{ShotcombiSQZ} have been divided by the measured shot noise. The almost white nonclassical noise suppression has continued up to the OPO cavity bandwidth of 27\,MHz. At frequencies below 10\,Hz the squeezing measurements suffered from non-stationary noise and the degree of squeezing could not be calibrated.
Peaks at 50\,Hz and 150\,Hz were picked from stray fields and are due to the electric mains.

\begin{figure}[t]
\hspace{-20mm}
\includegraphics[width=5.0cm]{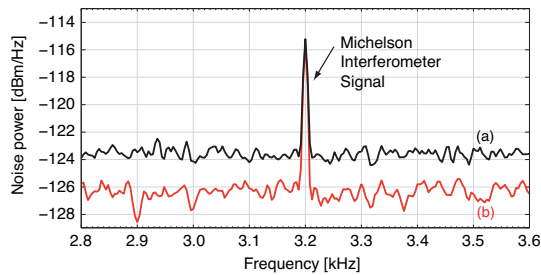}
  \vspace{0mm}
  \caption{(color online). Measured signal with quantum noise floors of a stably controlled Michelson Interferometer without (a) and with squeezed field injection (b). The latter one shows a nonclassical signal-to-noise improvement of 3 dB. The RBW was 4\,Hz, each point was averaged 200 times.}
  \label{Signal}
\end{figure}

In addition to demonstrating long term stable generation and observation of squeezed quantum noise in the gravitational wave band we applied the squeezed field for nonclassical sensitivity improvement of a Michelson Interferometer (MI) at audio frequencies. The squeezed field was injected into the interferometer's dark port via a Faraday rotator to replace the ordinary vacuum.  Since the MI was locked onto a dark fringe the squeezed quantum noise was reflected and interfered with the interferometer signal on the homodyne detector. A fringe visibility of 99.9\% at the 50/50 MI-beamsplitter was achieved.
The MI had an arm length of 40\,mm and was composed of two PZT mounted flat end mirrors (R=99.92\%). Interferometer laser power was 1.5$\mu$W. The error signal for locking the MI dark port was generated by longitudinal modulation of one end mirror at a frequency of 66\,kHz and proper demodulation of the MI bright port reflected light. 
The second PZT mounted MI-endmirror was modulated at 3.2\,kHz to simulate the effect of a gravitational-wave-signal. The angle of the squeezed quadrature was controlled to provide a nonclassical sensitivity improvement as shown in  Fig.\,\ref{Signal}.
In this experiment the squeezed beam experiences an extra 5\% transmission loss from double passing the rotator and additional 7\% loss due to a reduced homodyne fringe visibility (90.7\% instead of 94.3\%). This additional loss degraded the observed squeezing from about 4\,dB in  Fig.\,\ref{SQZdivSN} to then about 3\,dB.

In conclusion, we have reported on a new control scheme that allowed a stable generation, characterization and application of squeezed fields at audio frequencies. Broadband squeezing in the full gravitational wave detection band and a non-classical interferometer at audio frequencies have been demonstrated for the first time.  The coherent control technique reported in this paper provides an important stepping stone for the stable application of squeezed light in GW detectors.

This work has been supported by the Deutsche Forschungsgemeinschaft and is part of Sonderforschungsbereich 407.


\appendix 

\end{document}